\documentclass[aps,prl,twocolumn,superscriptaddress,showpacs,draft]{revtex4}
\usepackage{graphicx}
\input{epsf}

\begin{document}

\title{Google matrix and Ulam networks of intermittency maps}
\author{L.Ermann}
\affiliation{\mbox{Laboratoire de Physique Th\'eorique (IRSAMC), 
Universit\'e de Toulouse, UPS, F-31062 Toulouse, France}}
\affiliation{\mbox{LPT (IRSAMC), CNRS, F-31062 Toulouse, France}}
\author{D.L.Shepelyansky}
%\homepage[]{http://www.quantware.ups-tlse.fr}
\affiliation{\mbox{Laboratoire de Physique Th\'eorique (IRSAMC), 
Universit\'e de Toulouse, UPS, F-31062 Toulouse, France}}
\affiliation{\mbox{LPT (IRSAMC), CNRS, F-31062 Toulouse, France}}

%\date{\today}
\date{November  19, 2009 }

%\PACS{
%{05.45.Ac}{Low-dimensional chaos}
%\and
% 89.20.Hh  World Wide Web, Internet
%\and
%05.45.-a Nonlinear dynamics and chaos}

\begin{abstract}
We study the properties of the Google matrix
of an Ulam network generated by intermittency maps.
This network is created by the Ulam method which gives a
matrix approximant for the Perron-Frobenius operator of dynamical map.
The spectral properties of eigenvalues 
and eigenvectors of this matrix are analyzed.
We show that the PageRank of the system is 
characterized by a power law decay
with the exponent $\beta$ dependent on map parameters and 
the Google damping factor $\alpha$.  
Under certain conditions the PageRank is completely delocalized
so that the Google search in such a situation becomes inefficient.
\end{abstract}

\maketitle

\section{I Introduction}

In 60s Ulam proposed a method to construct a matrix approximant
for a Perron-Frobenius operator of dynamical systems which
is now known as the Ulam method \cite{ulam}.
The Ulam conjecture was that, in the limit of small 
cell discretization of the phase space, this method converges
and gives the correct description of the
 Perron-Frobenius operator of a system with
continuous phase space.
This conjecture was shown to be true for
hyperbolic maps of the interval
 \cite{li}. Various types of more generic maps of an interval
were studied in   \cite{tel,kaufmann,froyland2007}.
Further mathematical results have been obtained in 
\cite{ding,liverani,froyland2008a,froyland2008b} with extensions 
and prove of convergence for hyperbolic maps in
higher dimensions. The mathematical analysis of non-uniformly
expanding maps is now in progress \cite{murray2009}.
At the same time it is known that the Ulam method 
applied to Hamiltonian systems with integrable islands of motion
destroys the invariant curves thus producing a strong 
modification of properties of the Perron-Frobenius operator of
the system with continuous phase space (see e.g. \cite{dlszhirov}). 

Recently it was shown that the Ulam method naturally
generates a class of directed networks, named Ulam networks,
which properties have certain similarities with the
World Wide Web (WWW) networks \cite{dlszhirov}.  Thus the Google matrix
constructed for the Ulam networks built for the Chirikov 
typical map has a number of interesting
properties showing a power law decay of the PageRank vector.
  
The classification of network nodes by the PageRank Algorithm (PRA)
was proposed by Brin and Page in 1998 \cite{brin}
and became the core of the Google search engine
used everyday by majority of internet users.
The PRA is based on the construction of the Google matrix
which can be written as (see e.g. \cite{googlebook} for details):
\begin{equation}
{\bf G}=\alpha {\bf S}+(1-\alpha) {\bf E}/N \; .
\label{eq1}
\end{equation}
Here the matrix ${\bf S}$ is constructed from the adjacency matrix  ${\bf A}$
of  directed network links between $N$ nodes 
so that $S_{ij}=A_{ij}/\sum_k A_{kj}$ and
the elements of columns with
only zero elements are replaced by $1/N$. The  second term
in r.h.s. of (\ref{eq1}) describes  a finite probability $1-\alpha$
for WWW surfer to jump at random to any node so that the matrix elements
$E_{ij}=1$. This term stabilizes the convergence of PRA
introducing a gap between the maximal eigenvalue $\lambda=1$
and other eigenvalues $\lambda_i$. Usually the Google search uses the value
$\alpha=0.85$ \cite{googlebook}. The factor $\alpha$ is
also called the Google damping factor. By the construction 
$\sum_i G_{ij}=1$ so that the asymmetric matrix ${\bf G}$
belongs to the class of Perron-Frobenius operators.
Such operators naturally appear in the ergodic theory
\cite{sinai}
and dynamical systems
with Hamiltonian or dissipative dynamics \cite{mbrin,osipenko}.

The right eigenvector at $\lambda=1$ is 
the PageRank vector with 
positive elements $p_j$ and $\sum_j p_j=1$, 
the components $p_j$ of this vector 
are used for ordering and classification of nodes.
The PageRank can be efficiently obtained by a
multiplication of a random vector by ${\bf G}$
which is of low cost since in average there are
only about ten nonzero elements in a typical line
of ${\bf G}$ of WWW. This procedure converges
rapidly to the PageRank.
All WWW nodes can be ordered by decreasing 
$p_j$ ($p_{j} \geq p_{j+1}$)
so that the PageRank plays 
a significant role in the
ordering of websites and information retrieval.
The classification of nodes in the decreasing order of
$p_j$ values is used to classify
importance of network nodes as it is described in more detail in
\cite{googlebook}. 

Due to a spectacular success of the Google search
the studies of PageRank properties became very active
research filed  
in the computer science community.
A number of interesting results in this field can be find
in \cite{boldi,avrach1,litvak,avrach2}.
An overview of the field is available  in  \cite{avrach3}.
It is established that for large WWW subsets $p_j$
is satisfactory described by a scale-free algebraic decay
with $p_j \sim 1/j^{\beta}$ where $j$ is the PageRank ordering index
and $\beta \approx 0.9$ \cite{googlebook,donato}.

In this work we analyze the properties of Google matrix constructed
from Ulam networks generated by one-dimensional (1D) intermittency maps.
Such maps were introduced in \cite{pomeau} and studied in dynamical systems
with intermittency properties 
(see e.g. \cite{geisel1984,geisel1985,pikovsky1991,artuso2009}).
A number of mathematical results on the measure distribution
and slow mixing in such maps can be find in \cite{thaler1995,holland2005}
(see also Refs. therein). The 
mathematical properties of convergence of the Ulam method
in such intermittency maps are discussed in a recent work 
\cite{murray2009}. The analysis of such 1D maps
is simpler compared to the 2D map considered in \cite{dlszhirov}:
for example the PageRank at $\alpha=1$
is described by the invariant measure of the map
which can be find analytically as a function of map parameters. 
Following the approach discussed in \cite{dlszhirov,ggs}
we study not only the PageRank but also the spectrum
and the eigenstates of the Google matrix generated by the 
intermittency maps. 
Indeed, the right eigenvectors $\psi_i$ 
and eigenvalues $\lambda_i$ of the Google matrix
(${\bf G} \psi_i=\lambda_i \psi_i$)
are generally complex and
their properties should be studied in detail
to understand the behavior of the PageRank.
We show that under certain conditions the
properties of the PageRank can be drastically changed by 
parameter variation.

The results are presented in a following way: in  Section II
we describe the class of intermittency maps and the distribution of links
in the corresponding Ulam network, the spectral properties of the
Google matrix and PageRank are considered in Sections III and IV,
the discussion of the results is presented in Section IV.

\section{Intermittency maps}
%\vskip 0.3cm
\begin{figure}
\centerline{\epsfxsize=8.2cm\epsffile{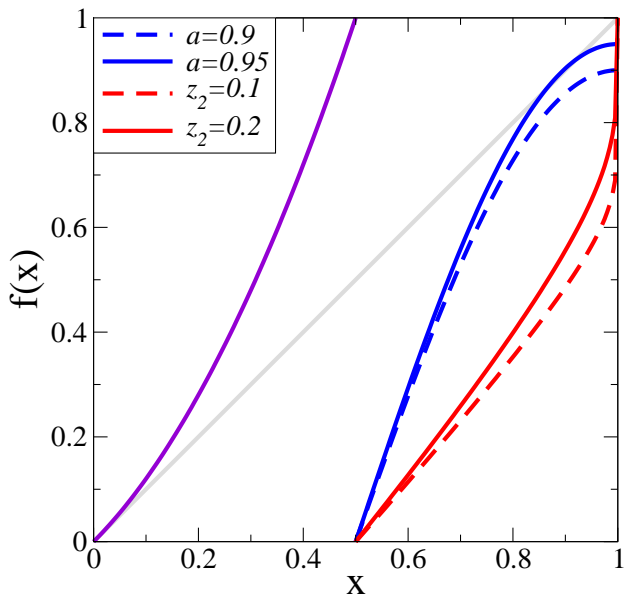}}
\vglue -0.2cm
\caption{(Color online) Two types of intermittency map of the interval
given by the map functions $f_1(x)$  and
$f_2(x)$, the functions are identical for 
$0 \leq x <1/2$ 
but have different branches at 
$1/2 \leq x \leq 1$ with $f_1(x)$ (red/gray) and $f_2(x)$ (blue/black); 
map functions are shown at $z_1=2$ and at different 
values of parameters $z_2$ and $a$; the straight line shows 
$f(x)=x$.
} 
\label{fig1}
\end{figure}

The intermittency maps of the interval considered in this paper
are described by the two map functions depending on parameters
and defined for the first model as:  

\begin{equation}
\label{eq2}
f_1(x)=
\left\{\begin{array}{ccc}
x+(2x)^{z_1}/2 \, , \, \text{for} \;\;\; 0\leq x<1/2\\
(2x-1-(1-x)^{z_2}+1/2^{z_2})/(1+1/2^{z_2}) \, , \, \\
      \text{for} \;\;\; 1/2\leq x \leq 1                    
\end{array} \right.
\end{equation}

and for the second model as

\begin{equation}
\label{eq3}
f_2(x)=
\left\{\begin{array}{ccc}
x+(2x)^{z_1}/2  \, , \, \text{for} \;\;\; 0\leq x<1/2\\
a\sin{\left[\pi(x-1/2)\right]} \, , \, \text{for} \;\;\; 1/2\leq x \leq1                    
\end{array} \right.
\end{equation}
\noindent
The parameters  $z_1, z_2, a$ are positive numbers. 
The dynamics is given by the map
$ \overline{x}=f_1(x)$ and $ \overline{x}=f_2(x)$.
The map functions $f_{1,2}(x)$ are shown in Fig.~\ref{fig1}. 
%\vskip 0.3cm
\begin{figure}
\centerline{\epsfxsize=7.1cm\epsffile{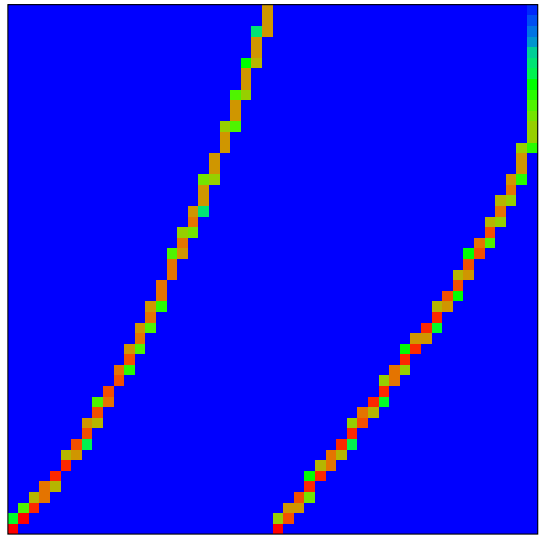}}
\vglue -0.2cm
\caption{(Color online) Google matrix at $\alpha=1$
generated by the intermittency map $f_1(x)$
at $z_1=2$, $z_2=0.2$, $N=50$, $N_c=10^6$
(amplitude of matrix elements is changing from zero (black/blue)
to 1 (red/gray)).
} 
\label{fig2}
\end{figure}

According to the usual theory of intermittency maps
and ergodicity theory 
\cite{pomeau,geisel1984,geisel1985,thaler1995,holland2005,ott,lichtenberg}
in the case of chaotic dynamics the steady state
invariant distribution $g(x)$ of the map is proportional to a time
$t(x)$ spent by a trajectory at point $x$ which is proportional
to $t \sim 1/x^{1-z_1}$ so that one has a power law distribution
at small values of $x$:
\begin{equation}
\label{eq4}
g(x) \propto 1/x^{z_1-1} \;\; .
\end{equation}
For $f_1$-map the dynamics is fully chaotic while for
$f_2$-map a fixed point attractor appears
for $a>0.945$ when $f_2(x)=x$.  

The Ulam networks generated by the intermittency maps 
(\ref{eq2}), (\ref{eq3}) are constructed in a way similar to one
described in \cite{ulam,dlszhirov}:
the whole interval $0\leq x \leq 1$ is divided on 
$N$ equal cells and $N_c$ trajectories (randomly distributed inside cell)
are iterated on one map iteration from cell $j$ to obtain matrix elements 
for transitions to cell $i$: $S_{ij} =N_i(j)/N_c$ where
$N_i(j)$ is a number of trajectories arrived from cell $j$ to cell $i$.
The image of the density of Google matrix elements is shown in Fig.~\ref{fig2}
for the first model.
The structure of the matrix repeats the form of the map
function $f_1(x)$.
We used from $10^4$ to $10^6$ cell trajectories $N_c$,
the obtained results are not sensitive to $N_c$
variation in this interval.
%\vskip 0.3cm
\begin{figure}
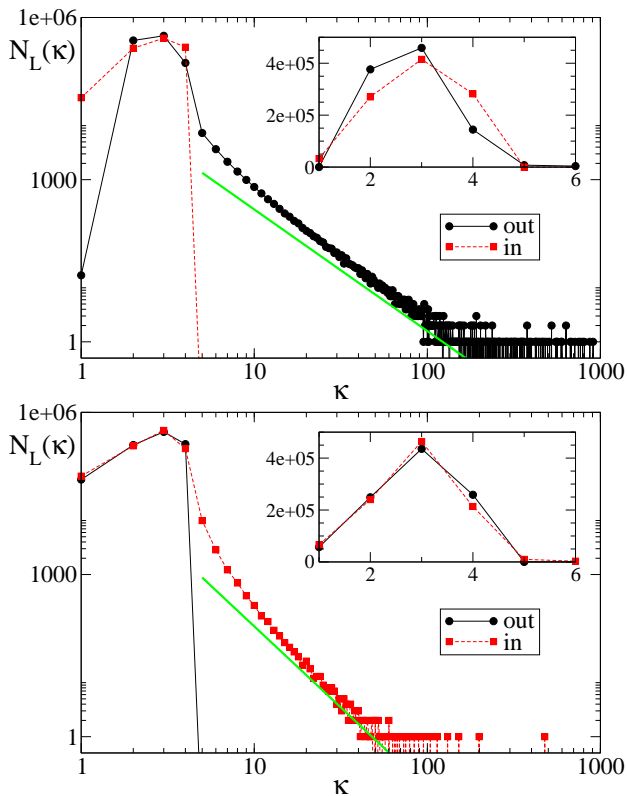

\centerline{\epsfxsize=8.2cm\epsffile{fig3a.eps}}
\vglue -0.0cm
\centerline{\epsfxsize=8.2cm\epsffile{fig3b.eps}}
\caption{(Color online) Distribution of links between nodes of the Ulam network
of $N=10^6$ size 
for the first model with $z_1=2, z_2=0.2$ (top panel) and the second model 
with $z_1=2, a=0.9$ (bottom panel). Here $N_L(\kappa)$ gives 
the number of nodes which have $\kappa$ outgoing (black points)
or ingoing (red/gray squares) links respectively. Insets show
data for small $\kappa$ values in linear scale. The straight line
shows the theoretical slope  for outgoing links
($N_L(\kappa) \propto 1/\kappa^{9/4}$ first model, top panel)
and for ingoing links ($N_L(\kappa) \propto 1/\kappa^{3}$ second model, 
bottom panel).
} 
\label{fig3}
\end{figure}

The differential distribution of number of nodes $N_L(\kappa)$ with 
ingoing or outgoing links $\kappa$ is shown in Fig.~\ref{fig3}. 
The first model shows  a sharp drop of 
ingoing links and a power law decay of
outgoing links. For the second model the situation is inverted.
These properties can be understood from the following arguments.
For the first model,
the number of outgoing links is $\kappa =d \overline{x}/dx = df_1(x)/dx$,
the derivative is diverging near $x =1$ where we have
$\kappa \sim 1/(1-x)^{(1-z_2)}$. The number of nodes
with $\kappa$ links is $N_n \sim (1-x) \sim 1/\kappa^{1/(1-z_2)}$
and the differential distribution of nodes
\begin{equation}
\label{eq5}
N_L^{out} \sim dN_n/dx \sim 1/\kappa^{\mu}, \; \mu=(2-z_2)/(1-z_2) \; .
\end{equation}
For the data of Fig.~\ref{fig3} (top panel) at $z_2=0.2$ this
estimate gives $\mu=9/4$ in a good agreement with the numerical data.
For the second model $df_2(x)/dx$ is always finite and we have a sharp
drop for outgoing links distribution.
The number of ingoing links is 
$\kappa = d x /d \overline{x} \sim 1/{\overline{x}}^{1-1/2\nu}$
since we have $\overline{x} \sim (1-x)^{2\nu}$ near $x=1$
(in our case $\nu=1$ but we consider here a general case).
Hence, the number of nodes with $\kappa$ links is $N_n \sim \overline{x} \sim
1/\kappa^{2\nu/(2\nu-1)}$ and 
\begin{equation}
\label{eq6}
N_L^{in} \sim dN_n/d\kappa \sim 1/\kappa^{\mu}, \; \mu=(4\nu-1)/(2\nu-1) \; . 
\end{equation}
For our case with $\nu=1$ we have
$\mu=3$.
This value is in a good agreement with the data of Fig.~\ref{fig3}.
For the first model $ d x /d \overline{x}$ is always finite and we have 
a sharp drop of ingoing links distribution.
 
This analysis allows to understand the origin
of power law distributions of links in the Ulam networks
generated by 1d maps.

\section{Spectral properties of the Google matrix}

%\vskip 0.3cm
\begin{figure}
\centerline{\epsfxsize=4.1cm\epsffile{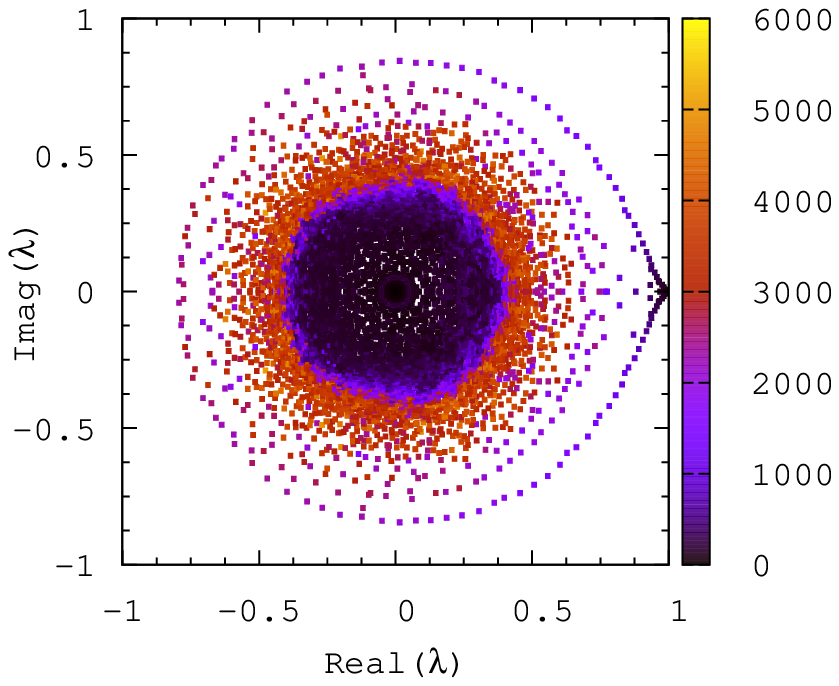}
\hfill\epsfxsize=4.1cm\epsffile{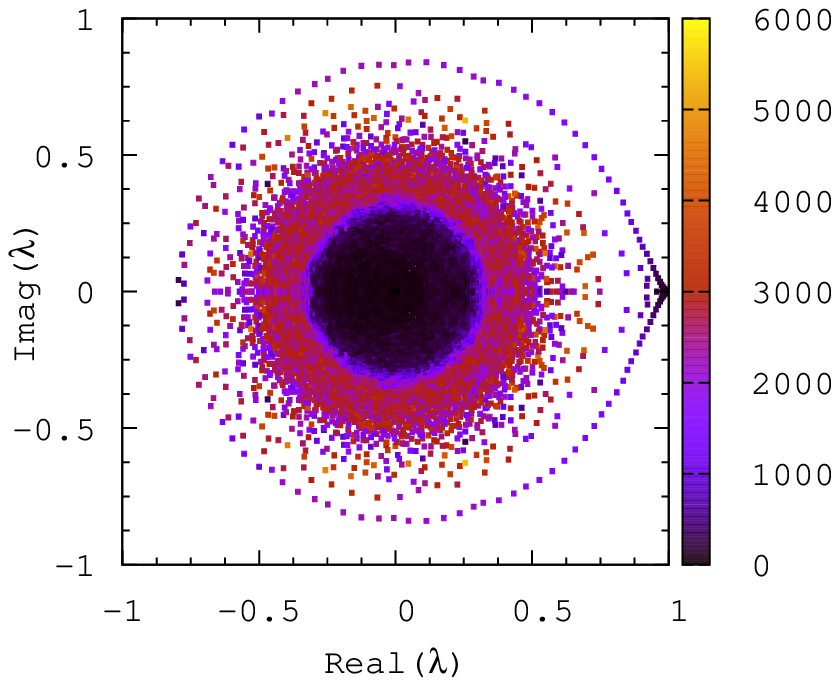}}
\vglue -0.2cm
\caption{(Color online) Distribution of eigenvalues $\lambda$ 
in the complex plain
for the Google matrix at  $\alpha=1$ for the first 
($z_1=2, z_2=0.2$, left panel)
and second ($z_1=2, a=0.9$, right panel)
models at $N=12000$. Color of small squares
is determined by the value of PAR $\xi$
associated with the corresponding eigenvector $\psi_i$ 
as show in the palette (the values of $\xi$
are averaged over the states inside of the square size).
} 
\label{fig4}
\end{figure}

The distribution of the eigenvalues of the Google matrix 
at $\alpha=1$ constructed
from the Ulam network described above is shown in Fig.~\ref{fig4}
for two models (\ref{eq2}) and (\ref{eq3}).
As in \cite{dlszhirov,ggs} we characterize an eigenstate $\psi_i$
by a PArticipation Ratio (PAR) defined
as $\xi_i= (\sum_j |\psi_i(j)|^2)^2/\sum_j |\psi_i(j)|^4$.
In fact PAR gives an effective number
of nodes populated by a given eigenstate, it is broadly used
in systems with disorder and Anderson localization.
The states $\psi_i(j)$ are normalized by
the condition $\sum_j |\psi_i(j)|^2=1$. For the PageRank
$p_j$  proportional to $\psi_1(j)$, ordered in the decreasing order
of probability, we use also probability normalization $\sum_j p_j=1$.

\begin{figure}
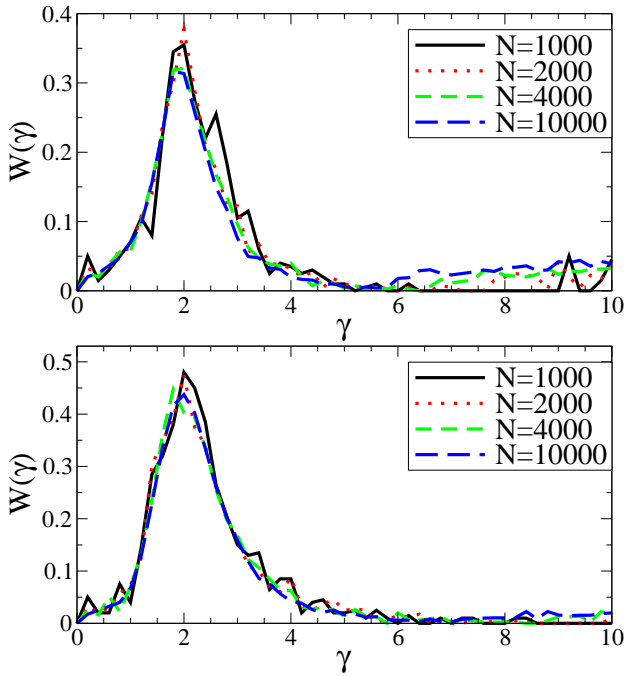

\centerline{\epsfxsize=8.2cm\epsffile{fig5a.eps}}
\vglue -0.0cm
\centerline{\epsfxsize=8.2cm\epsffile{fig5b.eps}}
\caption{(Color online)  Dependence of density of states 
$W(\gamma)$ on $\gamma$ shown for different values of $N$
for the first (at $z_1=2, z_2=0.2$, top panel)
and second (at $z_1=2, a=0.9$, bottom panel) models;
${\bf G}$ matrix is taken at $\alpha=1$.
} 
\label{fig5}
\end{figure}

There are few main features of the spectrum of $\lambda$ 
in Fig.~\ref{fig4} visible for two models:
there are states  with $|\lambda|$ close to 1 which
have relatively small values of $\xi$;
there is a circle like structure of eigenvalues 
and the maximum PAR are in the middle ring around the center. 
The large circle is present for both maps $f_1(x)$
and $f_2(x)$. This means that it appears due to the left branch
of the map corresponding to intermittent motion near $x =0$.
 The density distributions
$W(\gamma)=dN_\gamma/d\gamma$ 
in the decay rate defined as $\gamma=-2\ln|\lambda|$
are shown in Fig.~\ref{fig5}
(here $dN_\gamma$ is a number of states in the interval $d\gamma$). 
It is clear that in the limit of large
matrix size $N$ we have a convergence to a limiting distribution 
which has a characteristic peak at $\gamma \approx 2$.
%\vskip 0.3cm
\begin{figure}
\centerline{\epsfxsize=8.2cm\epsffile{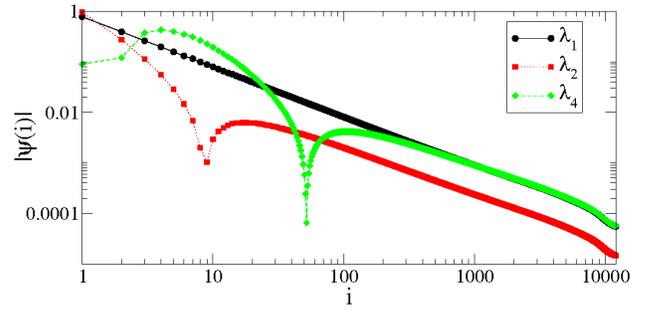}}
\vglue -0.2cm
\caption{(Color online) Absolute value of three eigenstates $\psi(i)$ for 
the first model $f_1(x)$ with $z_1=2$, $z_2=0.2$ and $N=12000$. 
The eigenstates correspond to the eigenvalues 
$\lambda_1=1.0$ (black circles), $\lambda_2 \approx 0.9998$  (red squares) 
and $\lambda_4\simeq0.9983$ (green diamonds) (see Fig.~\ref{fig4}, left panel). 
The corresponding PAR values are $\xi_1 \approx 2.54$, $\xi_2 \approx 1.21$ 
and $\xi_4 \approx 9.00$ respectively.
} 
\label{fig6}
\end{figure}

Examples of few eigenstates $\psi(i)$ with values of 
$\gamma_m=-2\ln|\lambda_m|$ equal and close to zero are shown
in Fig.~\ref{fig6} (the index $1 \leq i \leq N$ gives  
the cell position $x_i=(i-1)/N$,
index $m$ orders $\gamma_m$ from zero to maximum $\gamma$).
The first state $\psi_1(i)$ with $\lambda_1=1$
is the steady state distribution generated by the map $f_1(x)$
(the states for the map $f_2(x)$ have similar structure and 
we do not show them here).  We have $\psi_1(i) \propto 1/i^{\beta}$
with $\beta=1$ for $z_1=2$ is agreement with the
theoretical expression (\ref{eq4}) (the numerical fit gives $\beta=0.97$).
The state $\psi_1(i)$ is monotonic in $i$ so that
it coincides with the PageRank $p_j$ up to a constant factor. 
Eigenstates with next values of $\gamma$ are characterized by the same
decay at large $i$ with additional minima
at certain values of $i$ similar to few nodes of 
eigenstates in quantum mechanics.
%\vskip 0.3cm
\begin{figure}
\centerline{\epsfxsize=8.2cm\epsffile{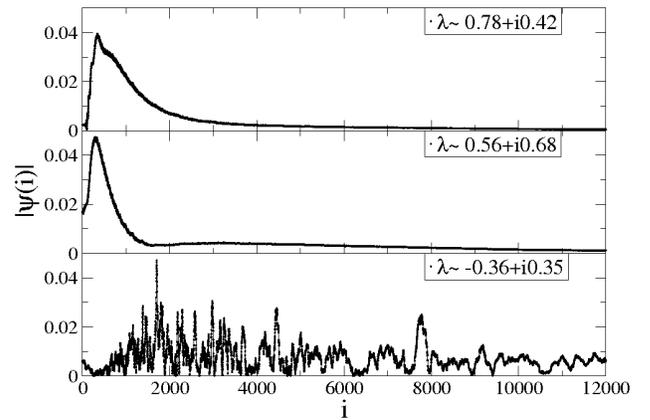}}
\vglue -0.2cm
\caption{Same as in Fig.~\ref{fig6} for the eigenstates
with $\lambda_{51}\approx 0.78 +i 0.42$ (a state in a large circle, top panel),
 $\lambda_{61}\approx 0.56 +i 0.68$ (a state in a large circle, middle panel)
and $\lambda_{1010} \approx -0.36 +i 0.35$ 
(a state in the dense part of the spectrum, bottom panel); 
the corresponding PARs are $\xi \approx 1231, 1482, 4367$
respectively.
} 
\label{fig7}
\end{figure}

The structure of eigenstates is changed when the value of $\gamma$ 
is increased. Typical states are shown in Fig.~\ref{fig7}.
The states on the first circle of $|\lambda|$  have peaked structure 
at certain $i$ with a plateau at large $i$. For $\gamma$ values
at the maximum of $W(\gamma)$ (see Fig.~\ref{fig5}) the eigenstates
are delocalized over the whole interval of $1 \leq i \leq N$.
%\vskip 0.3cm
\begin{figure}
\centerline{\epsfxsize=8.2cm\epsffile{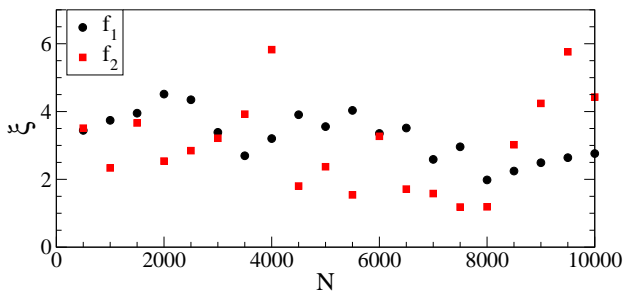}}
\vglue -0.2cm
\caption{Dependence of PAR $\xi$ of the PageRank on matrix size $N$:
the circles (black) are for the first model $f_1(x)$ with $z_1=2$, 
$z_2=0.2$ and the squares (red/grey) are for the second model $f_2(x)$
with $z_1=2$,  $a=0.9$.
} 
\label{fig8}
\end{figure}

\begin{figure}
\centerline{\epsfxsize=8.2cm\epsffile{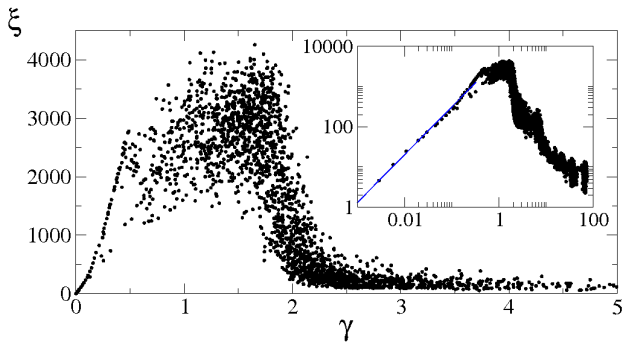}}
\vglue -0.2cm
\caption{Dependence of PAR $\xi$ on $\gamma$ 
 for the first model  with $z_1=2$,  
$z_2=0.2$ and $N=10^4$. Inset show data in log-log scale
with growth of $\xi \sim \gamma^{1.2}$ at small values of $\gamma$.
} 
\label{fig9}
\end{figure}

An effective number of sites contributing to an
eigenstate can be characterized by the PAR $\xi$.
For the PageRank the value of $\xi$ is independent of the matrix size $N$
as it is clearly shown in Fig.~\ref{fig8}. This is due to the power law
decay of the PageRank $p_j \sim 1/j$ which corresponds to 
an algebraic localization. The dependence of $\xi$ on $\gamma$
is shown in Fig.~\ref{fig9}. For small $\gamma$ it can be fitted by
a power law growth $\xi \sim \gamma^{1.2} $. The origin of the exponent
of this growth requires further analysis.

%\vskip 0.3cm

Finally we note that we also determined the dependence of 
number of states $N_\gamma$ with values of $\gamma>5$ on the
matrix size $N$. Our data (not shown) are well described by the dependence
$N_\gamma \sim N$ so that in contract to the results
presented in \cite{dlszhirov} there are no singes of the fractal Weyl law.
We attribute this to the fact that in contrast to the dissipative map
with a global contraction studied in \cite{dlszhirov} in the
intermittency maps all dynamics takes place on the whole 
one-dimensional interval with inhomogeneous distribution of measure 
but without fractality.

\section{Properties of the PageRank} 

\begin{figure}
\centerline{\epsfxsize=8.2cm\epsffile{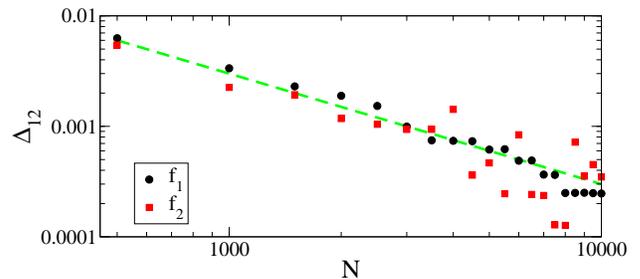}}
\vglue -0.2cm
\caption{Dependence of gap $\Delta_{12}=1-|\lambda_2|$ 
between the first eigenstate with
 $\lambda_1=1$ and next one with maximum $|\lambda_2|$ on $N$
 for the first $f_1(x)$ ($z_1=2$,  $z_2=0.2$) and second $f_2(x)$
($z_1=2$,  $a=0.9$) models. The straight dashed line shows the dependence
$\Delta_{12} \propto 1/N$.
} 
\label{fig10}
\end{figure}

The spectral gap between $\lambda_1=1$ equilibrium state
and the next state with maximum $|\lambda_2|$ has very small gap
$\Delta_{12}=1-|\lambda_2|$ which goes to zero with the increase of $N$
like $\Delta_{12} \approx 3/N$ (see Fig.~\ref{fig10}).
This happens due to the dynamical properties of the maps 
(\ref{eq2}), (\ref{eq3}) where the time spent at small $x \sim 1/N$
is of the order $t_x \sim 1/x^{z_1-1}$ 
(see e.g. \cite{pomeau,ott,lichtenberg}),
so that the corresponding
$ \Delta_{12} \sim 1/t_x \sim 1/N^{z_1-1}$ that gives 
the exponent $1$ for $z_1=2$.
%\vskip 0.3cm

\begin{figure}
\centerline{\epsfxsize=8.2cm\epsffile{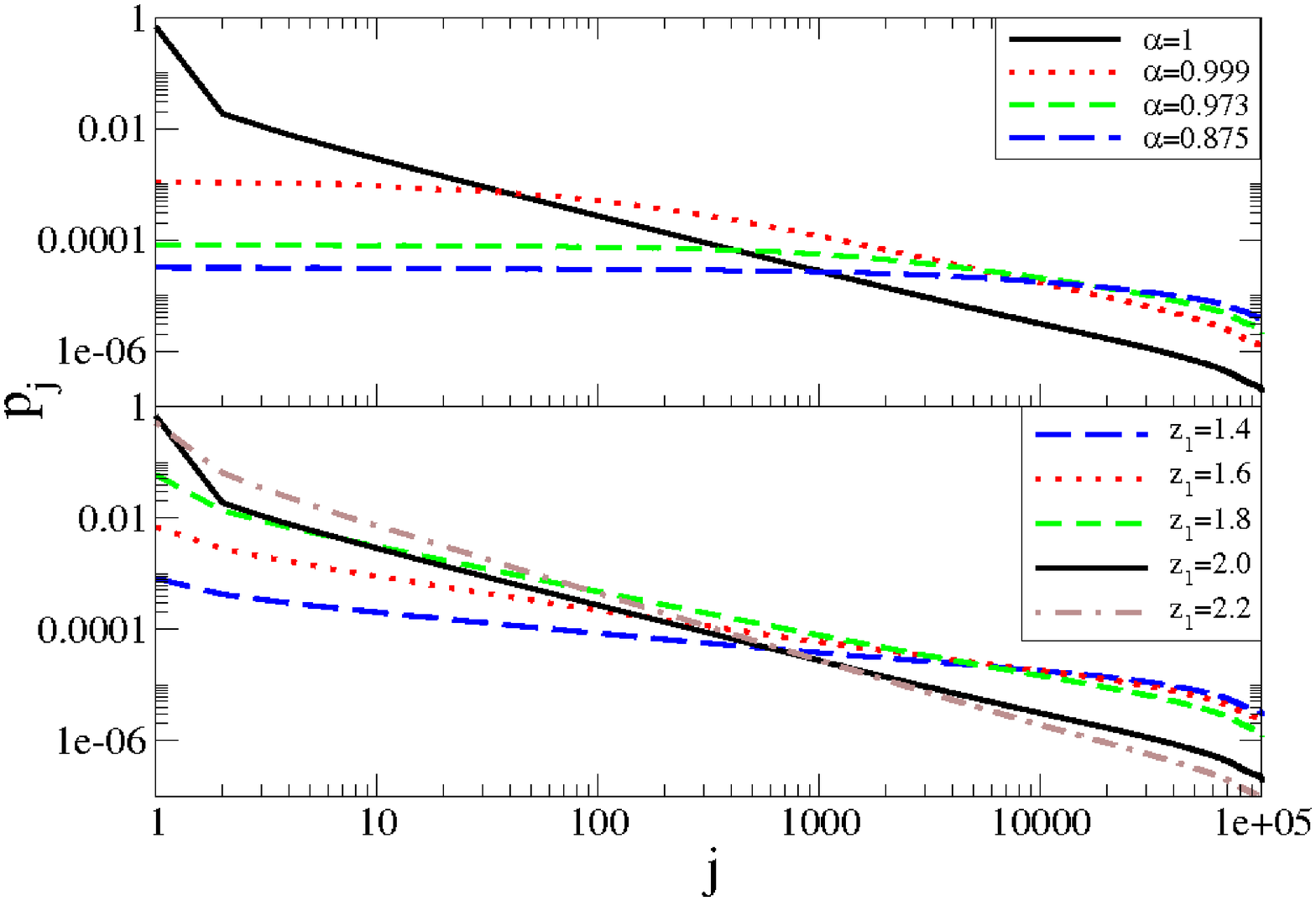}}
\vglue -0.0cm
\caption{(Color online) Dependence of the PageRank $p_j$ on $j$ for 
the first model at different values of $\alpha$ 
($z_1=2, z_2=0.2$, top panel)
and different values of $z_1$ ($z_2=0.2, \alpha=1$, bottom panel).
The data are obtained from the continuous map (see text) for $\alpha=1$
and the PageRank algorithm at $\alpha<1$, the number of nodes is $N=10^5$.
} 
\label{fig11}
\end{figure}

\begin{figure}
\centerline{\epsfxsize=8.2cm\epsffile{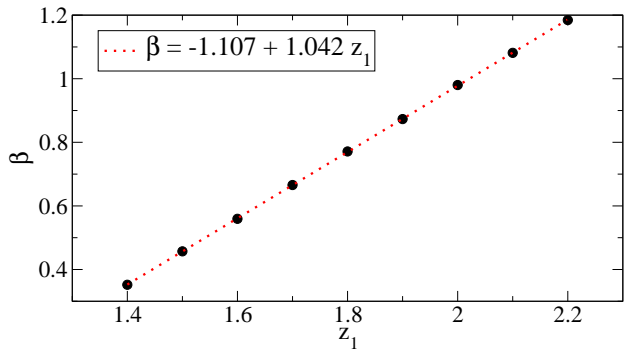}}
\vglue -0.2cm
\caption{Dependence of the PageRank exponent $\beta$ ($p_j \sim 1/j^\beta$)
on $z_1$ for the first model at $z_2=0.2$ and $\alpha=1$, $N=10^5$.
The straight dotted line shows the fit $\beta=1.042 z_1-1.107$.
} 
\label{fig12}
\end{figure}

Due to such decrease of $ \Delta_{12}$ with $N$ the PRA
has bad convergence at $\alpha=1$ for large values of $N$.
Up to $N\sim 14000$ we use the direct diagonalization of
$\bf{G}$ matrix which gives an algebraic decay
$p_j \sim 1/j^\beta$ with $\beta=1$ (see Fig.~\ref{fig6}).
For larger value of $N$ we used the continuous map
obtaining $p_j$ from an equilibrium distribution
over the cells of size $1/N$ after a larger
number of map iterations $t_i \approx 10^{9}$ and 
large number of trajectories
$N_{tr} \approx 10$. This distribution converges to a
limiting one at large values of $t_i$ (see Fig.~\ref{fig11}).
Both methods give the same result for $N<2 \cdot 10^4$.
The numerical data for the exponent $\beta$ 
are in good agreement with the theoretical dependence
(\ref{eq4}) $\beta=z_1-1$ as it is shown in Fig.~\ref{fig12}
(we attribute small deviations from the theoretical values to
finite size effects of $N$).
%\vskip 0.3cm

%\vskip 0.3cm

For $\alpha <1$ the PRA, described in the Introduction,
is stable and converges rapidly to the PageRank.
It gives the same results as the exact diagonalization for
$N< 2 \cdot 10^4$.
The dependence of PageRank on $\alpha$ 
is shown in Fig.~\ref{fig11} (top panel).
A small decrease down to $\alpha=0.999$ modifies $p_j$ at $j<100$
making $p_j$ very flat in this region. For $\alpha=0.875$
the PageRank becomes completely delocalized over the whole system size $N$.

%\vskip 0.3cm
\begin{figure}
\centerline{\epsfxsize=8.2cm\epsffile{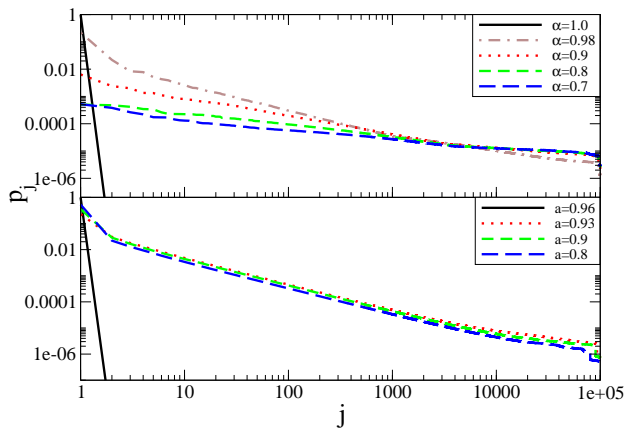}}
\vglue -0.0cm
\caption{(Color online) Dependence of the PageRank $p_j$ on $j$ for 
the second model at different values of $\alpha$ 
($z_1=2, a=0.96$, top panel)
and different values of $a$ ($z_1=2, \alpha=1$, bottom panel).
The data are obtained from the continuous map (see text) for $\alpha=1$
and the PageRank algorithm at $\alpha<1$, the number of nodes is $N=10^5$.
} 
\label{fig13}
\end{figure}
%\vskip 0.3cm
\begin{figure}
\centerline{\epsfxsize=4.4cm\epsffile{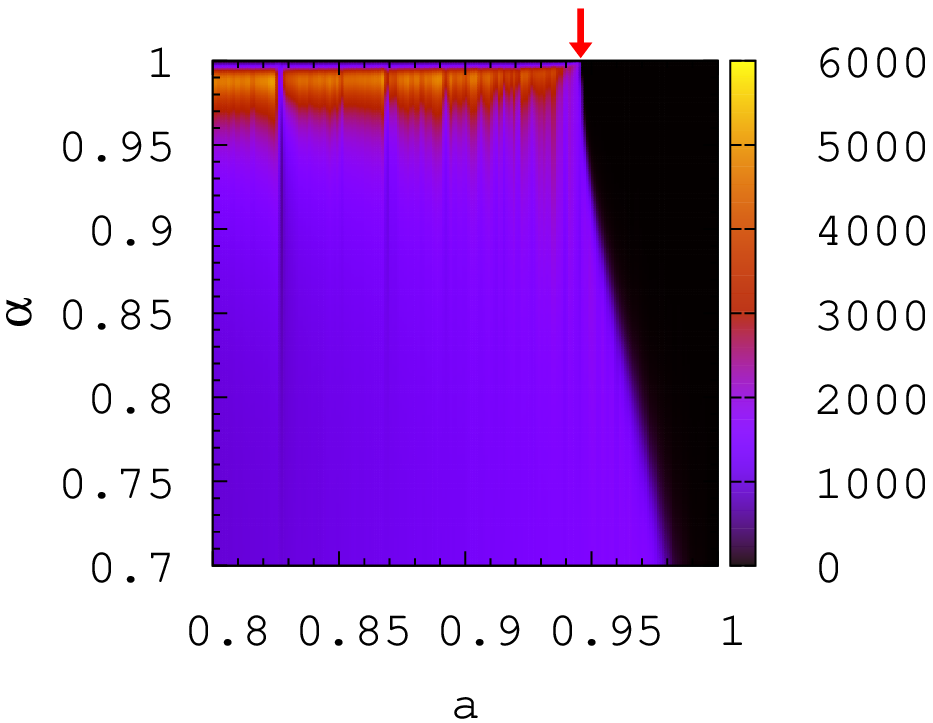}
\hfill\epsfxsize=4.4cm\epsffile{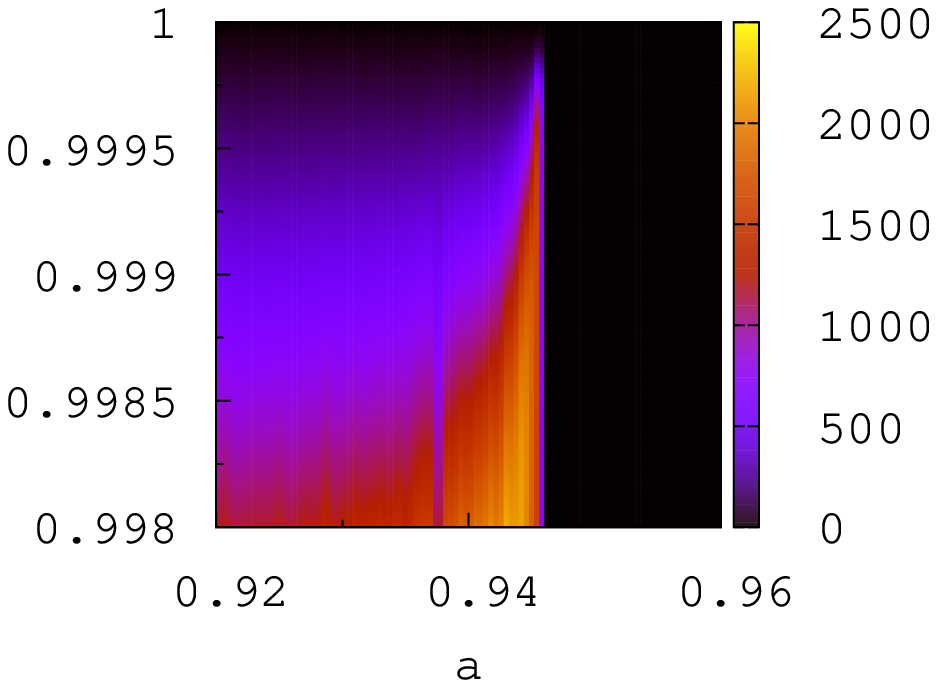}}
\vglue -0.2cm
\caption{(Color online) Dependence of PAR $\xi$ of the PageRank
($\xi$ values are shown by color) on parameters
$\alpha$ and $a$ for the second model at $z_1=2$; $N=10^5$;
arrow marks the region with higher resolution shown on the right panel. 
} 
\label{fig14}
\end{figure}

For the second model the PageRank depends strongly on the value of $a$.
For $a< 0.945$ when the dynamics is chaotic and the steady-state distribution 
is given by Eq.(\ref{eq4}) the properties of the PageRank are similar 
to those of the first model described above, e.g. we have $\beta =1$
being independent of $a$ for $\alpha=1, z_1=2$ 
(see Fig.~\ref{fig13}, bottom panel).
However, for $a>0.945$ the map has a fixed point attractor
and the Page Rank becomes localized practically on one site at $\alpha=1$.
In this regime with fixed point attractor the PageRank is very sensitive to
$\alpha$ variation: at $\alpha <1$ we have $p_j \sim 1/j^\beta$ 
with the fit values $\beta \approx 0.79$ at $\alpha=0.98$,
$\beta \approx 0.60$ at $\alpha=0.9$,   
$\beta \approx 0.42$ at $\alpha=0.8$ and $\beta \approx 0.32$ at $\alpha=0.7$.

The delocalization of the PageRank from the fixed point attractor state 
is also clearly seen in the variation of PAR $\xi(a,\alpha)$ shown 
in Fig.~\ref{fig14}. This shows that even if at $\alpha=1$
the PageRank is dominated only by one node a decrease of $\alpha$
allows to obtain weighted contribution of other nodes.

We also note that in the phase of fixed point attractor 
the spectrum of eigenvalues $\lambda$ has globally a structure
rather similar to one at $a=0.9 < 0.945$ (see Fig.~\ref{fig4}, right panel).
However, the PAR values of all eigenstates 
at $a> 0.945$ become rather close to unity
showing that almost all eigenstates are strongly localized in this  
phase. For example,  for $a=0.96$, we have 
almost all $\xi_i$ in the range from 1 to 4 for $N=10^4$,
it is interesting that about 53\% of the states have
$|\lambda| < e^{-10}$ (for $a=0.9$ this circle in $\lambda$
contains  23\% of states, see Fig.~\ref{fig4} right panel).

\section{Discussion}

The present studies allowed to establish a number of interesting properties
of the Google matrix constructed for the Ulam network generated by
intermittency maps. A general property of such networks
is the existence of states with eigenvalues $|\lambda|$ being very close to
unity. The PageRank of such networks at $\alpha=1$ is 
characterized by a power law decay with an exponent
determined by the parameters of the map.
It is interesting to note that usually for WWW it is
observed that the decay of the PageRank follows the decay law
of ingoing links distribution $N^{in}_L(\kappa)$ (see e.g. \cite{litvak}).
In our case the decay of PageRank is independent of 
$N^{in}_L(\kappa)$ decay as it is clearly shown by Eqs.~(\ref{eq5}),(\ref{eq6})
and the data of Figs.~\ref{fig3},\ref{fig11},\ref{fig13}.
In fact a map with singularities of both maps
$f_1(x)$ and $f_2(x)$ (e.g. $f_3(x)$ which behaves like
$x+x^{z_1}$ at small $x$, like $(1/2-x)^{z_1}$ at $x<1/2$ close to $1/2$
and like $(1-x)^\nu$ near $x=1$) will have 
the asymptotic decay of links distribution
given by Eqs.~(\ref{eq5}),(\ref{eq6}) but the decay of the PageRank will
be given by $\beta=z_1-1$, hence, being independent of the decay of
links distribution. 

Our results also show that while at $\alpha$ close to unity
the decay of the PageRank has the exponent $\beta \approx 1$
but at smaller values $\alpha \approx 0.9$
the PageRank becomes completely delocalized (see Fig.~\ref{fig11}).
In this delocalized phase the PAR $\xi$ grows
with the system size approximately as $\xi \propto N$.
The delocalization of the PageRank 
can also take place at $\alpha=1$ due to variation of the parameters of the map
(e.g. for $z_1 \rightarrow 1$). 
It is rather clear that the delocalization of 
the PageRank makes the Google search inefficient.

We hope that the properties of Ulam networks generated by simple maps
will be useful for future studies of real directed networks including WWW.
Indeed, the whole world will go blind if one day the Google search will become
inefficient. The investigations of the Ulam networks
can help to understand the properties of directed networks in a better way
that can help to prevent such a dangerous situation.

\section{Acknowledgments}

We thank A.S.Pikovsky for a useful discussion of his 
results presented in \cite{pikovsky1991}.

\end{document}